\documentclass[letterpaper, 11 pt]{article}

\usepackage{graphicx}                                                          
\usepackage{amsmath,amssymb}
\usepackage{amsthm}
\usepackage{amsmath}
\usepackage{amssymb}
\usepackage{xy}
\usepackage{array,hhline}
\usepackage{paralist}
\usepackage{tabularx}
\usepackage{tabulary}                                                           

\date{}

\title{\LARGE \bf
Robust Nonlinear $\mathcal{L}_2$ Filtering of
Uncertain Lipschitz Systems via Pareto Optimization}

\author{Masoud Abbaszadeh$^{\dag}$$^{\ddagger}$\thanks{Author to whom correspondence should be addressed.}\\
    masoud@ualberta.net
    \and
    Horacio J. Marquez$^\dag$\\
     marquez@ece.ualberta.ca
    }
\begin{document}
\maketitle \begin{center}\thanks{$^{\dag}$Department of Electrical
and Computer Engineering, University of Alberta, Edmonton, Alberta,
Canada, T6G 2V4 \\ $^{\ddagger}$Department of Research and Development, Maplesoft, Waterloo, Ontario, Canada, N2V 1K8}\end{center}

\begin{abstract}
A new approach for robust $H_{\infty}$ filtering for a class of
Lipschitz nonlinear systems with time-varying uncertainties both in the
linear and nonlinear parts of the system is proposed in an LMI framework.
The admissible Lipschitz constant of the system and the disturbance
attenuation level are maximized simultaneously through convex
multiobjective optimization. The resulting $H_{\infty}$ filter
guarantees asymptotic stability of the estimation error dynamics
with exponential convergence and is robust against nonlinear additive
uncertainty and time-varying parametric uncertainties.
Explicit bounds on the nonlinear uncertainty are derived
based on norm-wise and element-wise robustness analysis.
\end{abstract}

\emph{Keywords:} Nonlinear Uncertain Systems, Robust Observers,
Nonlinear $H_{\infty}$ Filtering, Convex Optimization

\section{Introduction}
The problem of observer design for nonlinear continuous-time
uncertain systems has been tackled in various approaches. Early
studies in this area go back to the works of de Souza \emph{et. al.}
where they considered a class of continuous-time Lipschitz nonlinear
systems with time-varying parametric uncertainties and obtained
Riccati-based sufficient conditions for the stability of the
proposed $H_{\infty}$ observer with guaranteed disturbance
attenuation level, when the Lipschitz constant is assumed to be
known and fixed, \cite{deSouza1}, \cite{deSouza2}. In an
$H_{\infty}$ observer, the $\mathcal{L}_{2}$-induced gain from the
norm-bounded exogenous disturbance signals to the observer error is
guaranteed to be below a prescribed level. They also derived matrix
inequalities helpful in solving this type of problems. Since then,
various methods have been reported in the literature to design
robust observers for nonlinear systems \cite{Nguang1,
McEneaney, Chen, Yung, Abbaszadeh1, Abbaszadeh4,abbaszadeh2008lmi,
abbaszadeh2010robust, abbaszadeh2012generalized,Pertew2, Xu3, Lu}. 
On the other hand, the restrictive
regularity assumptions in the Riccati approach can be relaxed using
linear matrix inequalities (LMIs). An LMI solution for nonlinear
$H_{\infty}$ filtering is proposed for Lipschitz nonlinear systems
with a given and fixed Lipschitz constant \cite{Xu3, Lu}. The
resulting observer is robust against time-varying parametric
uncertainties with guaranteed disturbance attenuation level.

In a recent paper the authors considered the nonlinear observer
design problem and presented a solution that has the following
features \cite{Abbaszadeh1}:
\begin{itemize}
\item (Stability)  In the absence of external disturbances the observer error converges to zero
exponentially with a guaranteed convergence rate. Moreover, our design is such that
it can maximize the size of the Lipschitz constant that can be tolerated in the system.
\item (Robustness) The design is robust with respect to
uncertainties in the nonlinear plant model.
\item (Filtering) The effect of exogenous disturbances on the observer error can be minimized.
\end{itemize}
In this article we consider a similar problem but consider the
important extension to the case where there exist parametric
uncertainties in the state space model of the plant. The extension
is significant because uncertainty in the state space model of the
plant is always encountered in a any actual application. Ignoring
this form of uncertainty requires lumping all model uncertainty on
the nonlinear (Lipschitz) term, thus resulting in excessively
conservative results. This extension, is though obtained through a
completely different solution from that given in \cite{Abbaszadeh1}.
The price of robustness against parametric uncertainties is an
stability requirement of the plant model which makes the solution,
different and yet a non-trivial extension to that of
\cite{Abbaszadeh1}. We will see this in detail in Section 3.
Our solution is based on the use of linear matrix inequalities and
has the property that the Lipschitz constant is one the LMI
variables. This property allows us to obtain a solution in which the
maximum admissible Lipschitz constant is maximized through convex
optimization. As we will see, this maximization adds an extra
important feature to the observer, making it robust against
nonlinear uncertainties.
The result is an $H_{\infty}$ observer
with a prespecified disturbance attenuation level which guarantees
asymptotic stability of the estimation error dynamics with
guaranteed speed of convergence and is robust against Lipschitz
nonlinear uncertainties as well as time-varying parametric
uncertainties, simultaneously. Explicit bound on the nonlinear
uncertainty are derived through a norm-wise analysis.
Some related results were recently presented by the authors in references
\cite{Abbaszadeh1} and \cite{Abbaszadeh4} for continues-time and for
discrete-time systems, respectively.
The rest of the paper is organized as follows.
In Section 2, the problem statement
and some preliminaries are mentioned. In Section 3, we propose a
new method for robust $H_{\infty}$ observer design for nonlinear
uncertain systems. Section 4, is devoted to robustness analysis in
which explicit bounds on the tolerable nonlinear uncertainty are
derived. In Section 5, a combined observer performance is optimized
using multiobjective optimization followed by a design example.

\section {Problem Statement}

Consider the following class of continuous-time uncertain nonlinear
systems:
\begin{align}
\left(\sum \right): \dot{x}(t)&=(A+\Delta A(t))x(t)+\Phi(x,u)+B
w(t)\label{sys1}
\\ y(t)&=(C+\Delta C(t))x(t)+Dw(t)\label{sys2}
\end{align}
where $x\in {\mathbb R} ^{n} ,u\in {\mathbb R} ^{m} ,y\in {\mathbb
R} ^{p} $ and $\Phi(x,u)$ contains nonlinearities of second order or
higher. We assume that the system (\ref{sys1})-\eqref{sys2} is
locally Lipschitz with respect to $x$ in a region $\mathcal{D}$
containing the origin, uniformly in $u$, i.e.:
\begin{align}
&\Phi(0,u^{*})=0\\
&\|\Phi(x_{1},u^{*})-\Phi(x_{2},u^{*})\|\leqslant\gamma\|x_{1}-x_{2}\|
\hspace{2mm}\forall \, x_{1},x_{2}\in \mathcal{D}
\end{align}
where $\|.\|$ is the induced 2-norm, $u^{*}$ is any admissible
control signal and $\gamma>0$ is called the Lipschitz constant. If
the nonlinear function $\Phi$ satisfies the Lipschitz continuity
condition globally in $\mathbb{R}^{n}$, then the results will be
valid globally. $w(t)\in\mathfrak{L}_{2}[0,\infty)$ is an unknown
exogenous disturbance, and $\Delta A(t)$ and $\Delta C(t)$ are
unknown matrices representing time-varying parameter uncertainties,
and are assumed to be of the form
\begin{eqnarray}
\Delta A(t)= M_{1}F(t)N_{1} \label{uncer1}\\
\Delta C(t)=M_{2}F(t)N_{2}\label{uncer2}
\end{eqnarray}
where $M_{1}$, $M_{2}$, $N_{1}$ are $N_{2}$ are known real constant
matrices and $F(t)$ is an unknown real-valued time-varying matrix
satisfying
\begin{equation}
F^{T}(t)F(t)\leq I \hspace{1cm} \forall t\in [0,\infty).
\end{equation}
The parameter uncertainty in the linear terms can be regarded as the
variation of the operating point of the nonlinear system. It is also
worth noting that the structure of parameter uncertainties in
(\ref{uncer1})-(\ref{uncer2}) has been widely used in the problems
of robust control and robust filtering for both continuous-time and
discrete-time systems and can capture the uncertainty in a number of
practical situations \cite{Khargonekar}, \cite{deSouza1},
\cite{Xu2}.

\subsection{Disturbance Attenuation Level}
Considering observer of the following form
\begin{eqnarray}
\dot{\hat{x}}(t)&=& A\hat{x}(t)+
\Phi(\hat{x},u)+L(y-C\hat{x})\label{observer1}
\end{eqnarray}
the observer error dynamics is given by
\begin{align}
e(t)\triangleq& \ x(t)-\hat{x}(t)\\
\begin{split}
\dot{e}(t)=& \ (A-LC)e+\Phi(x,u)-\Phi(\hat{x},u)\\&+(B-LD)w+(\Delta
A-L \Delta C)x.\label{error1}
\end{split}
\end{align}
Suppose that
\begin{equation}
z(t)=He(t)
\end{equation}
stands for the controlled output for error state where $H$ is a
known matrix. Our purpose is to design the observer parameter $L$
such that the observer error dynamics is asymptotically stable with
maximum admissible Lipschitz constant and the following specified
$H_{\infty}$ norm upper bound is simultaneously guaranteed.\
\begin{equation}
\|z\|\leq\mu\|w\|.
\end{equation}
Furthermore we want the observer to a have a guaranteed decay rate.

\subsection{Guaranteed Decay Rate}

Consider the nominal system $\left(\sum \right)$ with $\Delta
A,\Delta C=0$ and $w(t)=0$. Then, the decay rate of the system
(\ref{error1}) is defined to be the largest $\beta>0$ such that
\begin{eqnarray}
\lim_{t\rightarrow\infty} \exp(\beta t)\|e(t)\|=0
\end{eqnarray}
holds for all trajectories $e$. We can use the quadratic Lyapunov
function $V (e)=e^{T}Pe$ to establish a lower bound on the decay
rate of the (\ref{error1}). If $\frac{dV(e(t))}{dt}\leqslant-2\beta
V (e(t))$ for all trajectories, then $V(e(t)) \leqslant \exp(-2\beta
t)V(e(0))$, so that $\|e(t)\|\leqslant \exp(-\beta
t)\kappa(P)^{\frac{1}{2}}\|e(0)\|$ for all trajectories, where
$\kappa(P)$ is the condition number of P and therefore the decay
rate of the (\ref{error1}) is at least $\beta$, \cite{Boyd}. In
fact, decay rate is a measure of observer speed of convergence.

\section{$H_{\infty}$ Observer Synthesis}

In this section, an $H_{\infty}$ observer with guaranteed decay rate
$\beta$ and disturbance attenuation level $\mu$ is proposed. The
admissible Lipschitz constant is maximized through LMI optimization.
Theorem 1, introduces a design method for such an observer but first
we mention a lemma used in the proof of our result. It worths
mentioning that unlike the Riccati approach of \cite{deSouza1}, in
the LMI approach no regularity assumption is needed.\\

\emph{\textbf{Lemma 1. \cite{deSouza2}} Let $\mathcal{D}$,
$\mathcal{S}$ and $F$ be real matrices of appropriate dimensions and
$F$ satisfying $F^{T}F\leq I$. Then for any scalar $\epsilon>0$ and
vectors $x,y\in\mathbb{R}^{n}$, we have}
\begin{equation}
2x^{T} \mathcal{D}F\mathcal{S}y \leq
\epsilon^{-1}x^{T}\mathcal{D}\mathcal{D}^{T}x+\epsilon
y^{T}\mathcal{S}^{T}\mathcal{S}y
\end{equation}

\textbf{\emph{Note.}} As an standard notation in LMI context, the
symbol ``$\star$'' represents the element which makes the
corresponding matrix symmetric.\\

\emph{\textbf{Theorem 1.} Consider the Lipschitz nonlinear system
$\left(\sum \right)$ along with the observer \eqref{observer1}. The
observer error dynamics is (globally) asymptotically stable with
maximum admissible Lipschitz constant, $\gamma^{*}$, decay rate
$\beta$ and $\mathfrak{L}_{2}(w \rightarrow z)$ gain, $\mu$, if
there exists a fixed scalar $\beta>0$, scalars $\gamma>0$ and
$\mu>0$, and matrices $P_{1}>0$, $P_{2}>0$ and $G$, such
that the following LMI optimization problem has a solution.}\\
\begin{equation}
\hspace{-4cm} \max (\gamma) \notag
\end{equation}
\hspace{1cm} \emph{s.t.}
\begin{align}
&\left[
  \begin{array}{ccc}
    \Psi_{1} & 0 & \Omega_{1} \\
    \star & \Psi_{2} & \Omega_{2} \\
    \star & \star & -\mu^{2} I \\
  \end{array}
\right]<0 \label{LMI14}
\end{align}
where
\begin{eqnarray}
Q&=&-\left(A^{T}P_{1}+P_{1}A+2\beta P_{1}-C^{T}G^{T}-GC\right)\\
R&=&A^{T}P_{2}+P_{2}A+2N^{T}_{1}N_{1}+N^{T}_{2}N_{2}\label{R}\\
S&=&(I+M_{1}M_{1}^{T})^\frac{1}{2}\label{S}\\
 \Psi_{1}&=& \left[
  \begin{array}{cccc}
    H^{T}H-Q & \gamma I & P_{1}S & GM_{2} \\
    \star & -I & 0 & 0 \\
    \star & \star & -I & 0 \\
    \star & \star & \star & -I \\
  \end{array}
\right] \\
\Psi_{2}&=& \left[
\begin{array}{ccc}
  R & \gamma I & P_{2}S \\
  \star & -I & 0 \\
  \star & \star & -I
\end{array}
\right] \\
\Omega_{1}&=&\left[
             \begin{array}{cccc}
               P_{1}B-GD & 0 & 0 & 0 \\
             \end{array}
           \right]^{T}\\
\Omega_{2}&=&\left[
             \begin{array}{ccc}
               P_{2}B & 0 & 0 \\
             \end{array}
           \right]^{T}
\end{eqnarray}
\emph{Once the problem is solved}
\begin{eqnarray}
L&=&P_{1}^{-1}G \label{observer_gain}
\\\gamma^{*} &\triangleq& \max(\gamma)
\end{eqnarray}\\
\textbf{Proof:} From \eqref{error1}, the observer error dynamics is
\begin{equation}
\begin{split}
\dot{e}=& \ (A-LC)e+\Phi(x,u)-\Phi(\hat{x},u)+(B-LD)w\\&+(\Delta A-L
\Delta C)x.
\end{split}
\end{equation}
Let for simplicity
\begin{eqnarray}
\Phi(x,u)\triangleq \Phi,\ \ \Phi(\hat{x},u)\triangleq \hat{\Phi}.
\end{eqnarray}
Consider the Lyapunov function candidate
\begin{eqnarray}
V=V_{1}+V_{2}
\end{eqnarray}
where $V_{1}=e^{T}P_{1}e,\ \ V_{2}=x^{T}P_{2}x$.
For the nominal system, we have then
\begin{equation}
\begin{split}
\dot{V}_{1}(t)&=\dot{e}^{T}(t)P_{1}e(t)+e^{T}(t)P_{1}\dot{e}(t)\\&=-e^{T}Qe+2e^{T}P_{1}(\Phi(x,u)-\Phi(\hat{x},u))^{T}\label{V1}.
\end{split}
\end{equation}
To have $\dot{V}_{1}(t)\leqslant-2\beta V_{1}(t)$ it suffices
(\ref{V1}) to be less than zero, where:
\begin{equation}
(A-LC)^{T}P_{1}+P_{1}^{T}(A-LC)+2\beta P_{1}=-Q \label{lyap3}.
\end{equation}
The above can be written as
\begin{equation}
\\A^{T}P_{1}+P_{1}A-C^{T}L^{T}P_{1}-P_{1}LC+2\beta P_{1}=-Q.
\end{equation}
Defining the new variable
\begin{equation}
\\G\triangleq P_{1}L\Rightarrow L^{T}P_{1}^{T}=L^{T}P_{1}=G^{T},
\end{equation}
it becomes
\begin{equation}
\\A^{T}P_{1}+P_{1}A-C^{T}G^{T}-GC+2\beta P_{1}=-Q.
\end{equation}
Now, consider the systems $\left(\sum\right)$ with uncertainties and
disturbance. The derivative of $V$ along the trajectories of
$\left(\sum\right)$ is
\begin{equation}
\begin{split}
\dot{V}_{1}&=\dot{e}^{T}P_{1}e+e^{T}P_{1}\dot{e}\\
&=-e^{T}Qe+2e^{T}P_{1}(\Phi-\hat{\Phi})+2e^{T}P_{1}(B-LD)w\\
& \ \ \ +2e^{T}P_{1}M_{1}FN_{1}x-2e^{T}GM_{2}FN_{2}x.
\end{split}
\end{equation}
Using Lemma 1, it can be written
\begin{align}
&2e^{T}P_{1}M_{1}FN_{1}x\leq
e^{T}P_{1}M_{1}M^{T}_{1}P_{1}e+x^{T}N^{T}_{1}N_{1}x \label{ineq7} \\
&2e^{T}GM_{2}FN_{2}x\leq
e^{T}GM_{2}M^{T}_{2}G^{T}e+x^{T}N^{T}_{2}N_{2}x \label{ineq8}\\
&2x^{T}P_{2}M_{1}FN_{1}x\leq
x^{T}P_{2}M_{1}M^{T}_{1}P_{2}x+x^{T}N^{T}_{1}N_{1}x \label{ineq9}\\
&2e^{T}P_{1}(\Phi-\hat{\Phi})\leq
e^{T}P^{2}_{1}e+(\Phi-\hat{\Phi})^{T}(\Phi-\hat{\Phi})\notag \\
&\hspace{2.12cm}\leq e^{T}P^{2}_{1}e+\gamma^{2}e^{T}e \label{ineq10}\\
&2x^{T}P_{2}\Phi \leq x^{T}P^{2}_{2}x+\Phi^{T}\Phi \leq
x^{T}P^{2}_{2}x+{\gamma^{2}x^{T}x}\label{ineq11}
\end{align}
substituting from \eqref{ineq7}, \eqref{ineq8} and \eqref{ineq10}
\begin{equation}
\begin{split}
\dot{V}_{1}&\leq -e^{T}Q e+e^{T}P^{2}_{1}e+\gamma^{2}e^{T}e+e^{T}P_{1}M_{1}M^{T}_{1}P_{1}e\\
& \ \ \
+x^{T}(N^{T}_{1}N_{1}+N^{T}_{2}N_{2})x+e^{T}GM_{2}M^{T}_{2}G^{T}e
\\ & \ \ \ +2e^{T}P_{1}(B-LD)w.
\end{split}
\end{equation}
\begin{equation}
\begin{split}
\dot{V}_{2}&=x^{T}(A^{T}P_{2}+P_{2}A)x\\ & \ \ \ +2x^{T}P_{2}\Phi+2x^{T}P_{2}M_{1}FN_{1}x+2x^{T}P_{2}B w
\end{split}
\end{equation}
substituting from \eqref{ineq9}, \eqref{ineq11}
\begin{equation}
\begin{split}
\dot{V}_{2}&\leq
x^{T}(A^{T}P_{2}+P_{2}A)x+x^{T}P^{2}_{2}x+\gamma^{2}x^{T}x\\
& \
+x^{T}P_{2}M_{1}M^{T}_{1}P_{2}x+x^{T}N^{T}_{1}N_{1}x+2x^{T}P_{2}B w.
\end{split}
\end{equation}
Thus,
\begin{equation}
\begin{split}
\dot{V}&\leq
e^{T}\left[-Q+P_{1}(I+M_{1}M^{T}_{1})P_{1}+GM_{2}M^{T}_{2}G^{T}+\gamma
^{2} I\right]e \\
& \ \ +x^{T}\left[A^{T}P_{2}+P_{2}A+P_{2}(I+M_{1}M^{T}_{1})P_{2}+
\gamma
^{2}I\right]x\\
& \ \
+x^{T}(2N^{T}_{1}N_{1}+N^{T}_{2}N_{2})x+2e^{T}P_{1}(B-LD)w\\
& \ \ +2x^{T}P_{2}B w.\notag
\end{split}
\end{equation}
So, when $w=0$, a sufficient condition for the stability with
guaranteed decay rate $\beta$ is that
\begin{align}
&-Q+P_{1}SS^{T}P_{1}+GM_{2}M^{T}_{2}G^{T}+\gamma ^{2}I<0 \label{ineq12}\\
&R+P_{2}SS^{T}P_{2}+\gamma ^{2} I<0 \label{ineq13}
\end{align}
$R$ and $S$ are as in
\eqref{R} and \eqref{S}. Note that $I+M_{1}M^{T}_{1}$ is positive
definite and so has always a square root.
 Now, we define
\begin{equation}
J\triangleq \int^{\infty}_{0}(z^{T}z-\zeta w^{T}w) dt
\end{equation}
where $\zeta=\mu^{2}$. Therefore
\begin{equation}
J<\int^{\infty}_{0}(z^{T}z-\zeta w^{T}w+\dot{V}) dt
\end{equation}
so a sufficient condition for $J\leq0$ is that
\begin{equation}
\forall t\in[0,\infty),\hspace{5mm} z^{T}z-\zeta
w^{T}w+\dot{V}\leq0.
\end{equation}
We have
\begin{align}
z^{T}z-\zeta w^{T}w+\dot{V}\leq& e^{T}(H^{T}H-Q+P_{1}SS^{T}P_{1}
+GM_{2}M^{T}_{2}G^{T}+\gamma^{2} I)e\notag\\&+x^{T}(R+P_{2}SS^{T}P_{2}+\gamma^{2}I)x
+2e^{T}P_{1}(B-LD)w+2x^{T}P_{2}B w-\zeta w^{T}w \notag
\end{align}
So a sufficient condition for $J\leq0$ is that the right hand side
of the above inequity be less than zero which by means of Schur
complements is equivalent to (\ref{LMI14}). Note that \eqref{ineq12}
and \eqref{ineq13} are already included in \eqref{LMI14}. Then,
\begin{equation}
z^{T}z-\zeta w^{T}w\leq0\rightarrow\|z\|\leq\sqrt{\zeta}\|w\|. \ \
\blacksquare
\end{equation}

\textbf{Remark 1.} The proposed LMIs are linear in both $\gamma$ and
$\zeta(=\mu^{2})$. Thus, either can be a fixed constant or an
optimization variable. If one wants to design an observer for a
given system with known Lipschitz constant, then the LMI
optimization problem can be reduced to an LMI feasibility problem
(just satisfying the constraints) which is easier\\

\textbf{Remark 2.} This observer is  robust against two type of
uncertainties. Lipschitz nonlinear uncertainty in $\Phi(x,u)$ and
time-varying parametric uncertainty in the pair $(A,C)$ while the
disturbance attenuation level is guaranteed, simultaneously.

\section{Robustness Against Nonlinear Uncertainty}
As mentioned earlier, the maximization of Lipschitz constant makes
the proposed observer robust against some Lipschitz nonlinear
uncertainty. In this section this robustness feature is studied and
both norm-wise and element-wise bounds on the nonlinear uncertainty
are derived. The norm-wise analysis provides an upper bound on the
Lipschitz constant of the nonlinear uncertainty and the norm of the
Jacobian matrix of the corresponding nonlinear function.
Furthermore, we will find upper and lower bounds on the elements of
the Jacobian matrix through and element-wise analysis.

\subsection{Norm-Wise Analysis}
Assume a nonlinear uncertainty as follows
\begin{eqnarray}
\Phi_{\Delta}(x,u)&=&\Phi(x,u)+\Delta\Phi(x,u)
\\\dot{x}(t)&=& (A+ \Delta A)x(t) + \Phi_{\Delta}(x,u)
\label{uncer3}
\end{eqnarray}
where
\begin{eqnarray}
\|\Delta\Phi(x_{1},u)-\Delta\Phi(x_{2},u)\|\leqslant\Delta\gamma\|x_{1}-x_{2}\|.\\
\notag
\end{eqnarray}

\emph{\textbf{Proposition 1.}} {\emph{Suppose that the actual
Lipschitz constant of the system is $\gamma$ and the maximum
admissible Lipschitz constant achieved by Theorem 1, is
$\gamma^{*}$. Then, the observer designed based on Theorem 1, can
tolerate any additive Lipschitz nonlinear uncertainty with Lipschitz
constant less than or
equal $\gamma^{*}-\gamma$}}.\\

\textbf{Proof:} Based on Schwartz inequality, we have
\begin{eqnarray}
\begin{split}
\|\Phi_{\Delta}(x_{1},u)&-\Phi_{\Delta}(x_{2},u)\|\leq
\|\Phi(x_{1},u)-\Phi(x_{2},u)\|\\&\ \ \
+\|\Delta\Phi(x_{1},u)-\Delta\Phi(x_{2},u)\|
\\&\leq \gamma\|x_{1}-x_{2}\|+\Delta\gamma\|x_{1}-x_{2}\|.
\end{split}
\end{eqnarray}
According to the Theorem 1, $\Phi_{\Delta}(x,u)$ can be any
Lipschitz nonlinear function with Lipschitz constant less than or
equal to $\gamma^{*}$,
\begin{equation}
\|\Phi_{\Delta}(x_{1},u)-\Phi_{\Delta}(x_{2},u)\|\leq\gamma^{*}\|x_{1}-x_{2}\|
\end{equation}
so, there must be
\begin{eqnarray}
\gamma+\Delta\gamma\leq\gamma^{*}\rightarrow\Delta\gamma\leq\gamma^{*}-\gamma.\
\ \ \blacksquare
\end{eqnarray}
\\
 In addition, we know that for any continuously differentiable function $\Delta\Phi$,
\begin{eqnarray}
\|\Delta\Phi(x_{1},u)-\Delta\Phi(x_{2},u)\|\leqslant\|\frac{\partial\Delta\Phi}{\partial
x}(x_{1}-x_{2})\|
\end{eqnarray}
where $\frac{\partial\Delta\Phi}{\partial x}$ is the Jacobian matrix
\cite{Marquez}. So $\Delta\Phi(x,u)$ can be any additive uncertainty
with $\|\frac{\partial\Delta\Phi}{\partial x}\| \leq
\gamma^{*}-\gamma$. 

\subsection{Element-Wise Analysis}
Assume that there exists a matrix $\Gamma\in\mathbb{R}^{n\times n}$
such that
\begin{equation}
\|\Phi(x_{1},u)-\Phi(x_{2},u)\|\leqslant\|\Gamma(x_{1}-x_{2})\|.\label{Gamma}
\end{equation}
$\Gamma$ can be considered as a \emph{matrix-type Lipschitz
constant}. Suppose that the nonlinear uncertainty is as in
\eqref{uncer3} and
\begin{equation}
\|\Phi_{\Delta}(x_{1},u)-\Phi_{\Delta}(x_{2},u)\|\leqslant\|\Gamma_{\Delta}(x_{1}-x_{2})\|.\label{Gamma-Delta}
\end{equation}
Assuming
\begin{equation}
\|\Delta\Phi(x_{1},u)-\Delta\Phi(x_{2},u)\|\leqslant\|\Delta\Gamma(x_{1}-x_{2})\|\label{uncer4},
\end{equation}
based the proposition 1, $\Delta\Gamma$ can be any matrix with
$\|\Delta\Gamma\|\leq \gamma^{*}-\|\Gamma\|$. In the following, we
will look at the problem from a different angle. It is clear that
$\Gamma_{\Delta}=[{\gamma_{\Delta}}_{i,j}]_{n}$ is a perturbed
version of $\Gamma$ due to $\Delta \Phi(x,u)$. The question is that
how much perturbation can be tolerated on the element of $\Gamma$
without loosing the observer features stated in
Theorem 1. 
This is important in the sense that in gives us an insight about the
amount of uncertainty that can be tolerated in different directions
of the nonlinear function. Here, we propose a novel approach to
optimize the elements $\Gamma$ and provide specific upper and lower
bounds on tolerable perturbations. Before stating the result of this
section, we need to recall some matrix notations.

For matrices $A=[a_{i,j}]_{m\times n}$, $B=[b_{i,j}]_{m\times n}$,
$A\preceq B$ means $a_{i,j}\leq b_{i,j} \ \forall \ 1\leq i \leq m
,1\leq j\leq n$. For square A, $diag (A)$ is a vector containing the
elements on the main diagonal and $diag(x)$ where $x$ is a vector is
a diagonal matrix with the elements of $x$ on the main diagonal.
$|A|$
is the element-wise absolute value of $A$, i.e. $[|a_{i,j}|]_{n}$.
$A\circ B$ stands for the element-wise product (Hadamard product) of $A$ and $B$.\\

\emph{\textbf{Corollary 1.} Consider Lipschitz nonlinear system
$\left(\sum \right)$ satisfying \eqref{Gamma}, along with the
observer \eqref{observer1}. The observer error dynamics is
(globally) asymptotically stable with the matrix-type Lipschitz
constant $\Gamma^{*}=[\gamma^{*}_{i,j}]_{n}$ with maximized
admissible elements, decay rate $\beta$ and $\mathfrak{L}_{2}(w
\rightarrow z)$ gain, $\mu$, if there exist fixed scalars $\beta>0$
and $c_{i,j}>0 \ \forall \ 1\leq i,j\leq n$, scalars $\omega>0$ and
$\mu>0$, and matrices $\Gamma=[\gamma_{i,j}]_{n}\succ 0$, $P_{1}>0$,
$P_{2}>0$ and $G$, such that the following LMI optimization problem
has a solution.}
\begin{equation}
\hspace{-4cm} \max \ \omega \notag
\end{equation}
\hspace{0.8cm} \emph{s.t.}
\begin{align}
&c_{i,j}\gamma_{i,j}>\omega \ \ \ \ \ \ \forall \ 1\leq i,j\leq n\\
&\left[
  \begin{array}{ccc}
    \Psi_{1} & 0 & \Omega_{1} \\
    \star & \Psi_{2} & \Omega_{2} \\
    \star & \star & -\mu^{2} I \\
  \end{array}
\right]<0
\end{align}
\emph{where $\Psi_{1}$, $\Psi_{2}$, $\Omega_{1}$ and $\Omega_{2}$
are as in Theorem 1 replacing $\gamma I$ by $\Gamma$. Once the
problem is solved}
\begin{eqnarray}
L&=&P_{1}^{-1}G
\\\gamma^{*}_{i,j} &\triangleq& \max(\gamma_{i,j})
\end{eqnarray}\\
\textbf{Proof:} The proof is similar to the proof of Theorem 1 with
replacing $\gamma I$ by $\Gamma$. \ \ \ $\blacksquare$\\

\textbf{Remark 3.} By appropriate selection of the weights
$c_{i,j}$, it is possible to put more emphasis on the directions in
which the tolerance against nonlinear uncertainty is more important.
To this goal, one can take advantage of the knowledge
about the structure of the nonlinear function $\Phi(x,u)$.\\

According to the norm-wise analysis, it is clear that $\Delta\Gamma$
in \eqref{uncer4} can be any matrix with $\|\Delta\Gamma\|\leq
\|\Gamma^{*}\|-\|\Gamma\|$. We will now
proceed by deriving bounds on the elements of $\Gamma_{\Delta}$.\\

\emph{\textbf{Lemma 2.} For any $T=[t_{i,j}]_{n}$ and
$U=[u_{i,j}]_{n}$, if $|T|\preceq U$, then $TT^{T}\leq UU^{T}\circ
nI$}.\\


\textbf{Proof:} Assume any $x=[x_{i}]_{n\times1}$, then, it is easy
to show that $T^{T}x=[(\sum_{i=1}^{n}t_{i,j}x_{i})_{j}]_{n\times1}$.
Therefore,
\begin{equation}
\begin{split}
x^{T}TT^{T}x&=\langle T^{T}x,T^{T}x\rangle=\sum_{j=1}^{n}(\sum_{i=1}^{n}t_{i,j}x_{i})^{2}\\
&\leq\sum_{j=1}^{n}\sum_{i=1}^{n}t_{i,j}^{2}x_{i}^{2}+\sum_{j=1}^{n}\sum_{i=1}^{n-1}\sum_{k=i+1}^{n}(t_{i,j}^{2}x_{i}^{2}+t_{k,j}^{2}x_{k}^{2})\\
&\leq\sum_{j=1}^{n}\sum_{i=1}^{n}u_{i,j}^{2}x_{i}^{2}+\sum_{j=1}^{n}\sum_{i=1}^{n-1}\sum_{k=i+1}^{n}(u_{i,j}^{2}x_{i}^{2}+u_{k,j}^{2}x_{k}^{2})\\
&=\sum_{j=1}^{n}[\sum_{i=1}^{n}u_{i,j}^{2}x_{i}^{2}+\sum_{i=1}^{n-1}\sum_{k=i+1}^{n}(u_{i,j}^{2}x_{i}^{2}+u_{k,j}^{2}x_{k}^{2})]\\
&=n\sum_{i=1}^{n}\sum_{j=1}^{n}u_{i,j}^{2}x_{i}^{2}=n \ x^{T}diag(diag(UU^{T}))x\\
&\Rightarrow \ TT^{T}\leq n \ diag(diag(UU^{T}))= UU^{T}\circ nI.\notag \ \blacksquare
\end{split}
\end{equation}
Now we are ready to state the element-wise robustness result. Assume
additive uncertainty in the form of \eqref{uncer3}, where
\begin{equation}
\|\Phi_{\Delta}(x_{1},u)-\Phi_{\Delta}(x_{2},u)\|\leqslant\|\Gamma_{\Delta}(x_{1}-x_{2})\|.\label{Gamma-Delta}
\end{equation}
It is clear that $\Gamma_{\Delta}=[{\gamma_{\Delta}}_{i,j}]_{n}$ is
a perturbed version of $\Gamma$.\\

\emph{\textbf{Proposition 2.}} \emph{Suppose that the actual
matrix-type Lipschitz constant of the system is $\Gamma$ and the
maximized admissible matrix-type Lipschitz constant achieved by
Corollary 1, is $\Gamma^{*}$. Then, $\Delta\Phi$ can be any additive
nonlinear uncertainty such that $|\Gamma_{\Delta}|\preceq
n^{-\frac{3}{4}} \Gamma^{*}$.}\\

\textbf{Proof:} According to the Proposition 1, it suffices to show
that $\sigma_{max}(\Gamma_{\Delta})\leq\sigma_{max}(\Gamma^{*})$.
Using Lemma 2, we have
\begin{equation}
\begin{split}
\sigma_{max}^{2}(\Gamma_{\Delta})&=\lambda_{max}(\Gamma_{\Delta}\Gamma_{\Delta}^{T})\\
&\leq\lambda_{max}(n\ diag(diag(n^{-\frac{3}{2}}\Gamma^{*}{\Gamma^{*}}^{T})))\\
&\leq \sigma_{max}(n\ diag(diag(n^{-\frac{3}{2}} \Gamma^{*}{\Gamma^{*}}^{T})))\\
&=\max_{i}(n^{-\frac{1}{2}}\sum_{j=1}^{n}{\gamma^{*}_{i,j}}^{2})=\frac{1}{\sqrt{n}}\|\Gamma^{*}\circ\Gamma^{*}\|_{\infty}\\
&\leq|\Gamma^{*}\circ\Gamma^{*}\|_{2}\leq
|\Gamma^{*}\|^{2}_{2}=\sigma_{max}^{2}(\Gamma^{*}).\notag
\end{split}
\end{equation}
The first inequality follows from Lemma 2 and the symmetry of
$\Gamma_{\Delta}\Gamma_{\Delta}^{T}$ and
diag(diag($\Gamma^{*}{\Gamma^{*}}^{T}))$, \cite{Horn1}. The last two
inequalities are due to the relation between the induced infinity
and 2 norms \cite{Horn1} and the fact that the spectral norm is
submultiplicative with respect to the Hadamard product \cite{Horn2},
respectively. Since the singular values are nonnegative, we can
conclude that
$\sigma_{max}(\Gamma_{\Delta})\leq\sigma_{max}(\Gamma^{*})$. \ \ \
$\blacksquare$\\

Therefore, denoting the elements of $\Gamma_{\Delta}$ as
${\gamma_{\Delta}}_{i.j}=\gamma_{i,j}+\delta_{i,j}$, the following
bound on the element-wise perturbations is obtained

\begin{equation}
-n^{-\frac{3}{4}}\gamma^{*}_{i,j}-\gamma_{i,j} \leq \delta_{i,j}
\leq n^{-\frac{3}{4}}\gamma^{*}_{i,j}-\gamma_{i,j}.
\end{equation}

In addition, $\Delta\Phi(x,u)$ can be any continuously
differentiable additive uncertainty which makes
$|\frac{\partial\Phi_{\Delta}}{\partial x}|\preceq n^{-\frac{3}{4}}
\Gamma^{*}$. It is worth mentioning that the results of Lemma 2 and
Proposition 2 have intrinsic importance from the matrix analysis
point of view regardless of our specific application in the
robustness analysis.

\section{Combined Performance using Multiobjective Optimization}

The LMIs proposed in Theorem 1 are linear in both admissible
Lipschitz constant and disturbance attenuation level. So, as
mentioned earlier, each can be optimized. A more realistic problem
is to choose the observer gain matrix by combining these two
performance measures. This leads to a Pareto multiobjective
optimization in which the optimal point is a trade-off between two
or more linearly combined optimality criterions. Having a fixed
decay rate, the optimization is over $\gamma$ (maximization) and
$\mu$ (minimization), simultaneously. The following theorem is in
fact a generalization of the results of \cite{Xu3} and \cite{Xu4}
(for the systems in class of $\sum$) in which the Lipschitz constant
is known and fixed, in one point of view; and the results of
\cite{Howell} in which a special class of sector nonlinearities
is considered and there is no uncertainty in pair (A,C), in another.\\

\emph{\textbf{Theorem 2.} Consider Lipschitz nonlinear system
$\left(\sum \right)$ along with the observer \eqref{observer1}. The
observer error dynamics is (globally) asymptotically stable with
decay rate $\beta$ and simultaneously maximized admissible Lipschitz
constant, $\gamma^{*}$ and minimized $\mathfrak{L}_{2}(w \rightarrow
z)$ gain, $\mu^{*}$, if there exists fixed scalars $\beta>0$ and
$0\leq\lambda\leq1$, scalars $\gamma>0$ and $\zeta>0$, and matrices
$P_{1}>0$, $P_{2}>0$ and $G$, such
that the following LMI optimization problem has a solution.}\\
\begin{equation}
\hspace{-4cm} \min \left[\lambda(-\gamma)+(1-\lambda)\zeta \right]
\notag
\end{equation}
\hspace{0.8cm} \emph{s.t.}
\begin{align}
&\left[
  \begin{array}{ccc}
    \Psi_{1} & 0 & \Omega_{1} \\
    \star & \Psi_{2} & \Omega_{2} \\
    \star & \star & -\zeta I \\
  \end{array}
\right]<0
\end{align}
\emph{where $\Psi_{1}$, $\Psi_{2}$, $\Omega_{1}$ and $\Omega_{2}$
are as in Theorem 1. Once the problem is solved}
\begin{eqnarray}
L&=&P_{1}^{-1}G
\\\gamma^{*} &\triangleq& \max(\gamma)=\min(-\gamma)
\\\mu^{*} &\triangleq& \min(\mu)=\sqrt{\zeta}
\end{eqnarray}\\
\textbf{Proof:} The above is a scalarization of a multiobjective
optimization with two optimality criterions. Since each of these
optimization problems is convex, the scalarized problem is also
convex \cite{Boyd2}. The rest of the proof is the
same as the proof of Theorem 1. $\blacksquare$ \\

\textbf{Remark 4.} The matrix-type Lipschitz constant $\Gamma$ may
also be considered in place of $\gamma$ in Theorem 2.\\

Since the observer gain directly amplifies the measurement noise,
sometimes, it is better to have an observer gain with smaller
elements. There might also be practical difficulties in implementing
high gains. We can control the Frobenius norm of $L$ either by
changing the feasibility radius of the LMI solver or by decreasing
$\lambda_{min}^{-1}(P_{1})$ which is $\lambda_{max}(P_{1}^{-1})$, to
decrease $\bar{\sigma}(L)$  as in (\ref{observer_gain}). The latter
can be done by replacing $P_{1}>0$ with $P_{1}>\theta I$ in which
$\theta>0$ can be either a fixed scalar or an LMI variable.
Considering $\bar{\sigma}(L)$ as another performance index, note
that it is even possible to have a triply combined cost function in
the LMI optimization problem of Theorem 2. Now, we show the
usefulness of this
Theorem through a design example.\\

\begin{figure}[!h]
  \centering
  \includegraphics[width=4in]{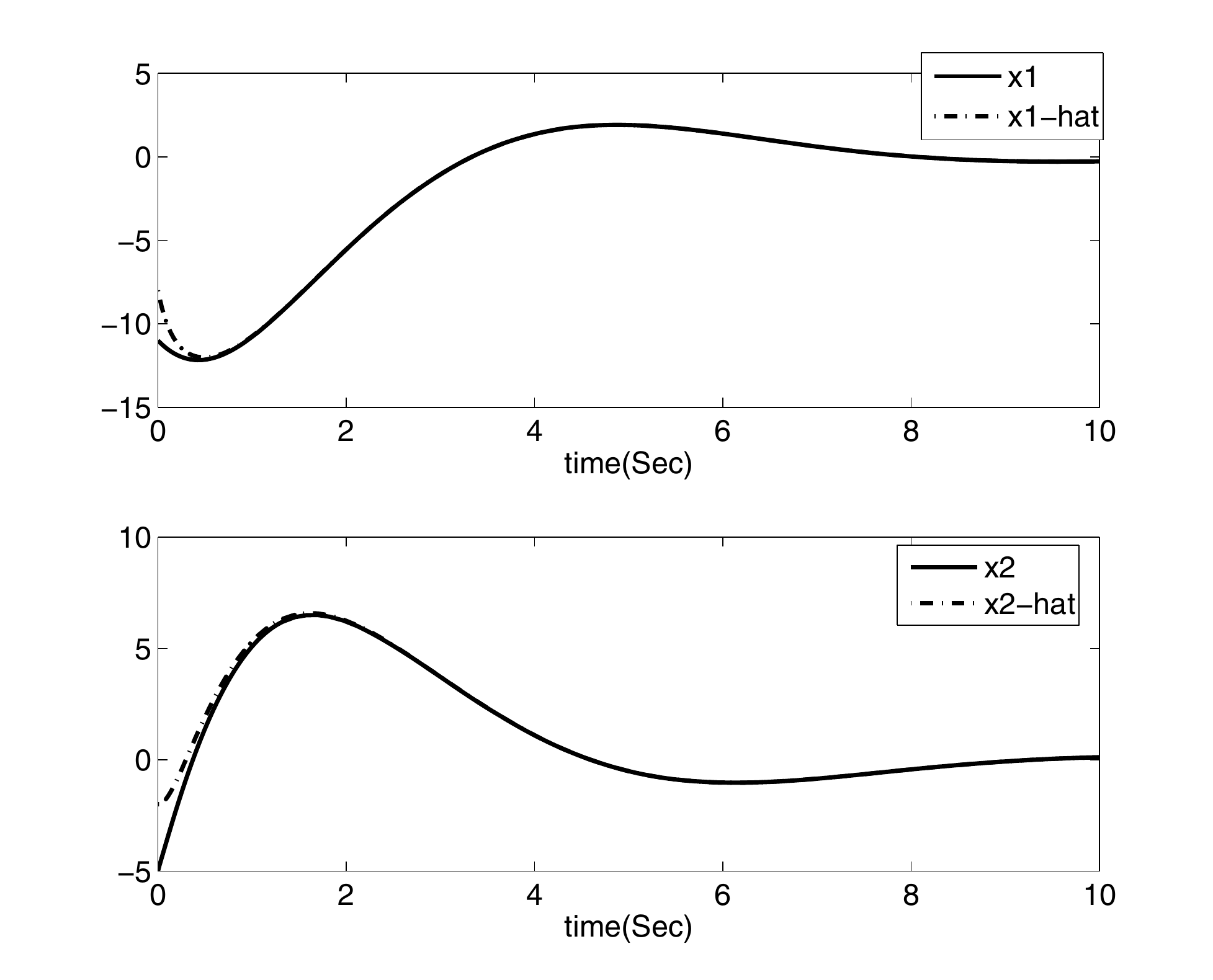}\\
  \caption{The true and estimated states of the example}\label{Fig1}
\end{figure}
\begin{figure}[!h]
  \centering
  \includegraphics[width=4in]{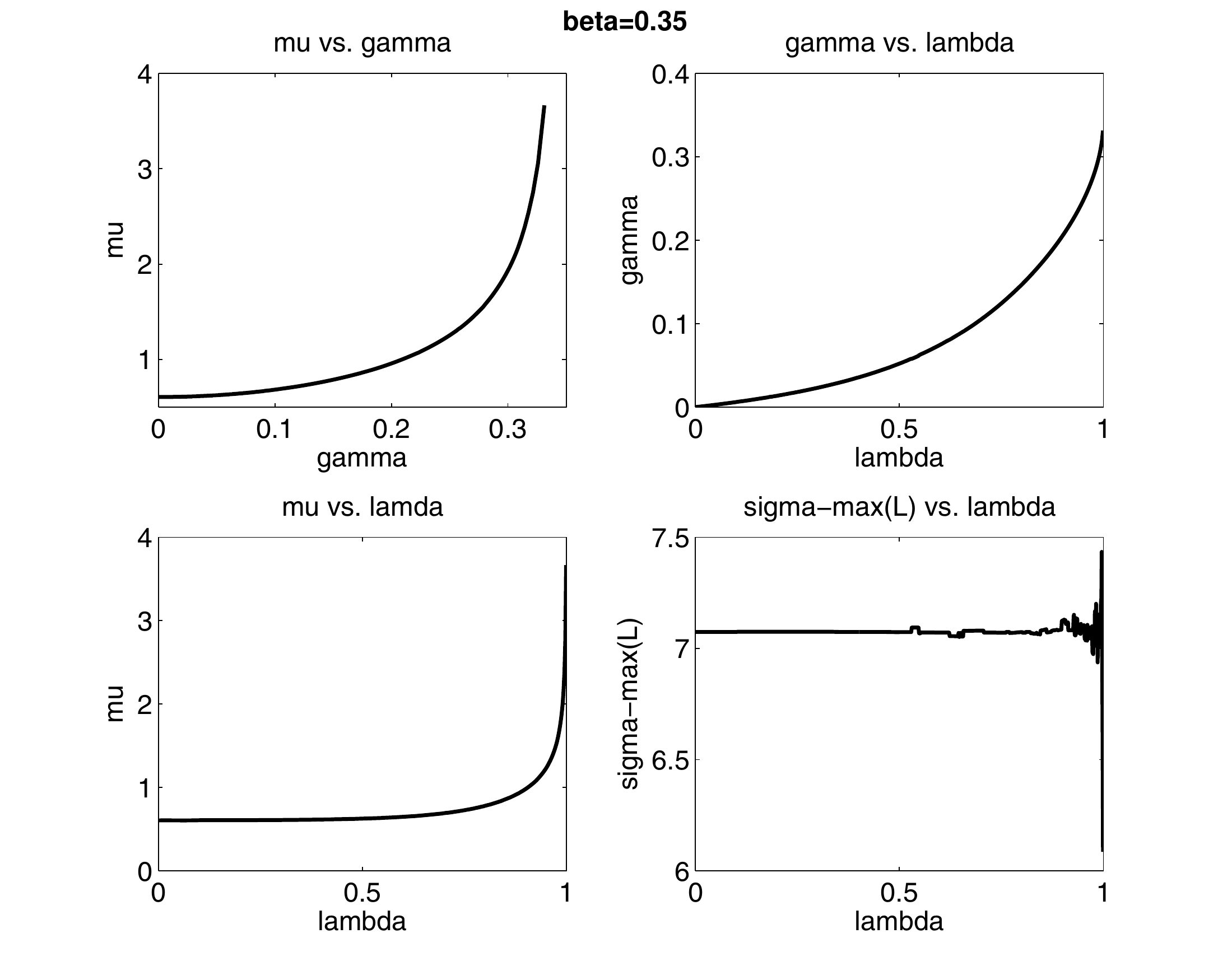}\\
  \caption{$\gamma^{*}$, $\mu^{*}$ and
$\bar{\sigma}(L)$, and the optimal trade-off curve}\label{Fig2}
\end{figure}
\begin{figure}[!h]
  \centering
  \includegraphics[width=4in]{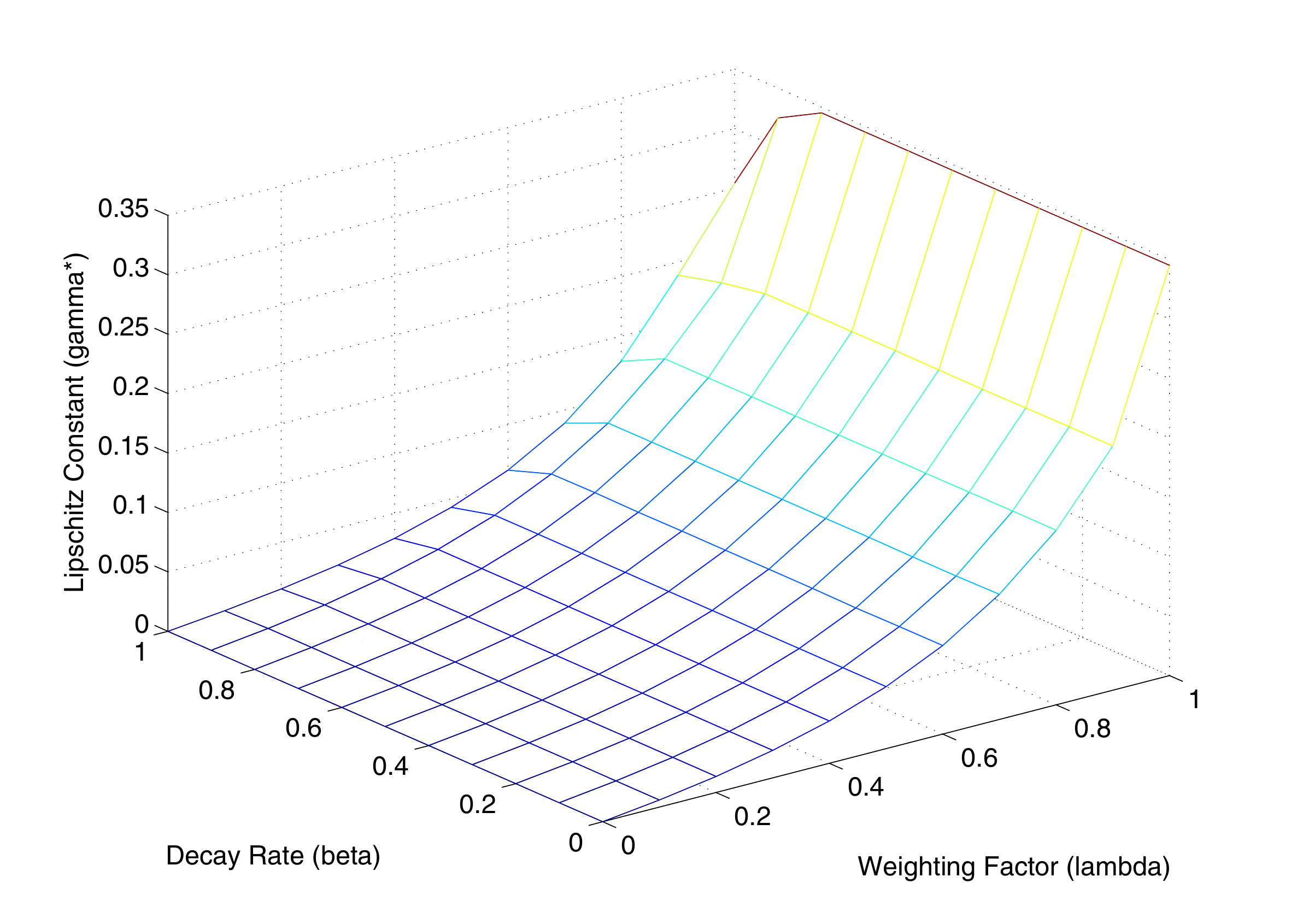}\\
  \caption{The optimal surface of $\gamma^{*}$}\label{Fig3}
\end{figure}
\begin{figure}[!h]
  \centering
  \includegraphics[width=4in]{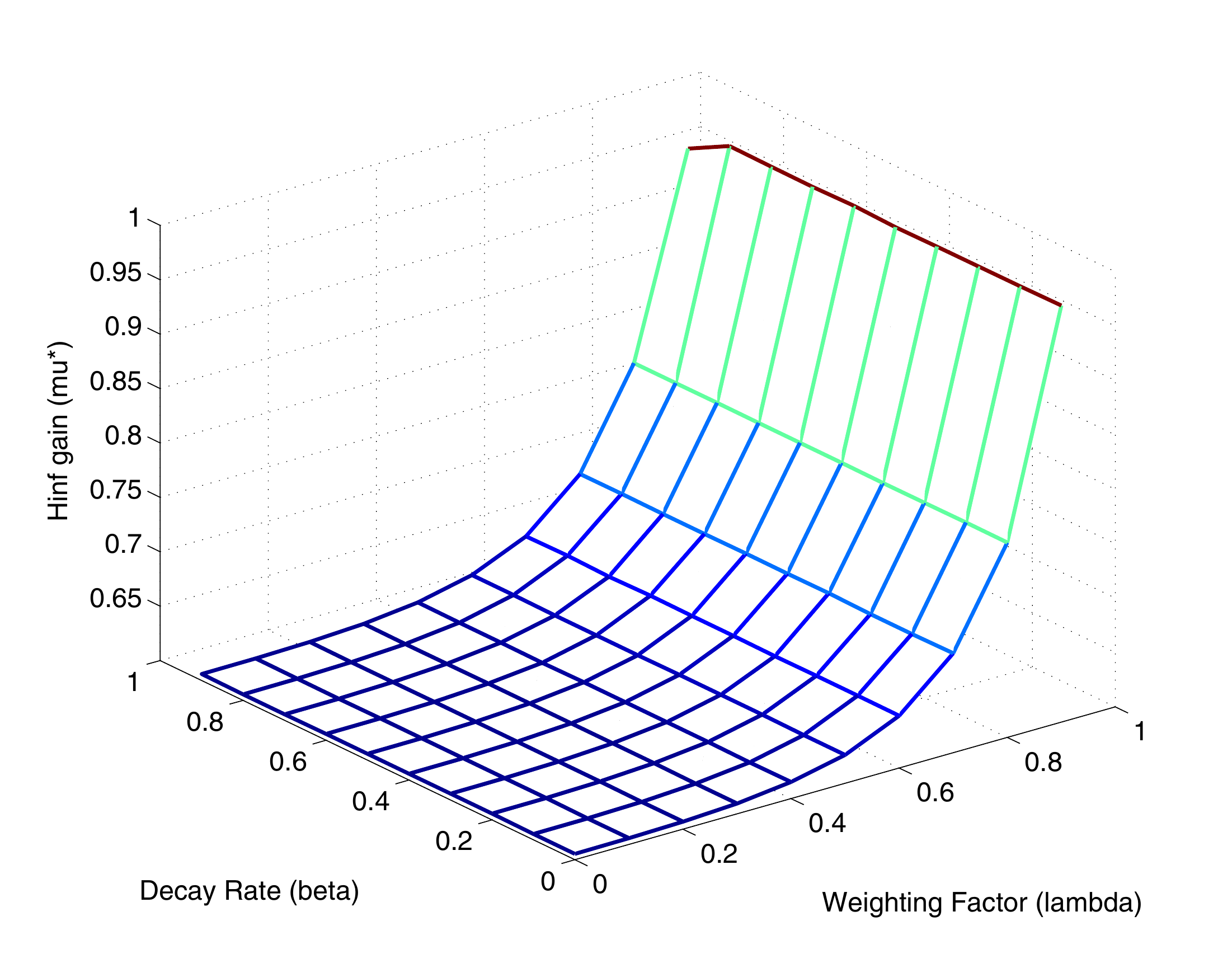}\\
  \caption{The optimal surface of $\mu^{*}$}\label{Fig4}
\end{figure}
\begin{figure}[!h]
  \centering
  \includegraphics[width=4in]{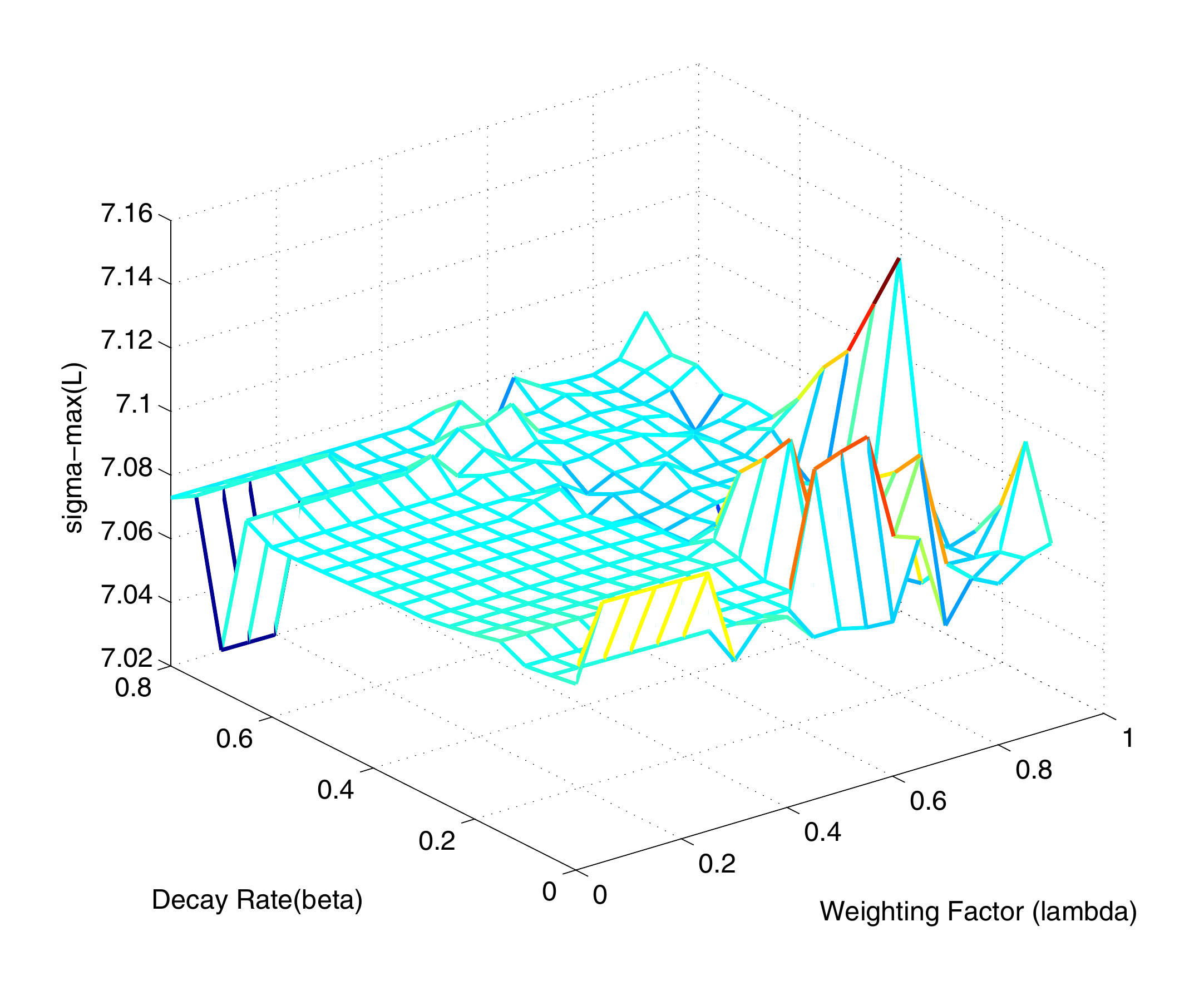}\\
  \caption{The optimal surface of $\bar{\sigma}(L)$}\label{Fig5}
\end{figure}

\textbf{Example:} Consider a system of the form of
$\left(\sum\right)$ where \label{example4}
\begin{eqnarray}
A&=&\left[
    \begin{array}{cc}
      0 & 1 \\
      -1 & -1 \\
    \end{array}
  \right], \ \
\Phi(x)=\left[
          \begin{array}{c}
            0 \\
            0.2sin(x_{1}) \\
          \end{array}
        \right]\notag\\
M_{1}&=&\left[
  \begin{array}{cc}
    0.1 & 0.05 \\
    -2 & 0.1 \\
  \end{array}
\right], \ \ M_{2}=\left[
     \begin{array}{cc}
       -0.2 & 0.8 \\
     \end{array}
   \right]\notag\\
C&=&\left[
    \begin{array}{cc}
      1 & 0 \\
    \end{array}
  \right], \ \ N_{1}=N_{2}=\left[
              \begin{array}{cc}
                0.1 & 0 \\
                0 & 0.1 \\
              \end{array}
            \right].\notag
\end{eqnarray}
Assuming
\begin{eqnarray}
\beta&=&0.35, \lambda=0.95\notag\\
B&=&\left[
      \begin{array}{cc}
        1 & 1 \\
      \end{array}
    \right]^{T}\notag\\
D&=&0.2\notag\\
H&=&0.5 I_{2}\notag
\end{eqnarray}
we get
\begin{eqnarray}
\gamma^{*}&=&0.3016, \ \mu^{*}=3.5\notag\\
L&=&\left[
        \begin{array}{cc}
          5.0498 & 4.9486 \\
        \end{array}
      \right]^{T}\notag
\end{eqnarray}
Figure \ref{Fig1}, shows the true and estimated values of states.
 The values of $\gamma^{*}$, $\mu^{*}$ and
$\bar{\sigma}(L)$, and the optimal trade-off curve between
$\gamma^{*}$ and $\mu^{*}$ over the range of $\lambda$ when the
decay rate is fixed ($\beta=0.35$) are shown in figure \ref{Fig2}.
The optimal surfaces of $\gamma^{*}$, $\mu^{*}$ and
$\bar{\sigma}(L)$ over the range of $\lambda$ when the decay rate is
variable are shown in figures \ref{Fig3}, \ref{Fig4} and \ref{Fig5},
respectively.
The maximum value of $\gamma^{*}$ is 0.34 obtained
when $\lambda=1$. 
In the range of $0\leq\lambda\leq1$ and $0\leq\beta\leq0.8$, the
norm of $L$ is almost constant. As $\beta$ increases over 0.8,
$\bar{\sigma}(L)$ rapidly increases and for $\beta=1.2$, the LMIs
are infeasible.

\section{Conclusion}

A new nonlinear $H_{\infty}$ observer design method for a class of
Lipschitz nonlinear uncertain systems is proposed through LMI
optimization. The developed LMIs are linear both in the admissible
Lipschitz constant and the disturbance attenuation level allowing
both two be an LMI optimization variable. The combined performance
of the two optimality criterions is optimized using Pareto
optimization. The achieved $H_{\infty}$ observer guarantees
asymptotic stability of the error dynamics with a prespecified decay
rate (exponential convergence) and is robust against Lipschitz
additive nonlinear uncertainty as well as time-varying parametric
uncertainty. Explicit bounds on the nonlinear uncertainty are
derived through norm-wise and element-wise analysis.


\bibliographystyle{plain}
\bibliography{Candidacy_References}


\end{document}